\documentstyle[aps,prb,preprint,tighten,graphicx]{revtex}
\begin{document}
\draft

\title{Antiresonances in Molecular Wires}

\author{Eldon G. Emberly\footnote{e-mail: eemberly@sfu.ca  copyright IOP 
1999} and George Kirczenow}

\address{Department of Physics, Simon Fraser University,
Burnaby, B.C., Canada V5A 1S6}

\date{\today}

\maketitle
\begin{abstract}
We present analytic and numerical studies based on Landauer
theory of conductance antiresonances of molecular wires.  Our
analytic treatment is a solution of the Lippmann-Schwinger
equation for the wire that includes the effects of the
non-orthogonality of the atomic orbitals on different atoms
exactly. The problem of non-orthogonality is treated by
solving the transport problem in a new Hilbert space which is
spanned by an orthogonal basis. An expression is derived for
the energies at which antiresonances should occur for a
molecular wire connected to a pair of single-channel 1D leads. 
From this 
expression we identify two distinct mechanisms that
give rise to antiresonances under different circumstances.
The exact treatment of non-orthogonality in the theory is
found to be necessary to obtain reliable results. Our
numerical simulations extend this work to multichannel leads
and to molecular wires connected to 3D metallic
nanocontacts. They demonstrate that our analytic results also
provide a good description of these more complicated systems
provided that certain well-defined conditions are met. These
calculations suggest that antiresonances should be
experimentally observable in the differential conductance of
molecular wires of certain types.

\end{abstract}
\section{Introduction}
There has been renewed interest recently in molecular
wires,\cite{conf} stimulated in part by experimental work that
has begun to explore possible ways of measuring the conductance
of a single molecule.\cite{Reed97,Datta97,Andres96,Bumm96}
Theoretically, electron transport in molecular wires has been
studied by considering the transmission probability for
electrons to scatter through the
structure.\cite{Datta97,Samant96,Kemp96,Joach96,Mujic97,Ember98}
As with other mesoscopic systems, the electrical conductance
$G$ of the molecule is related to the transmission probability
$T$ at the Fermi level by the Landauer formula\cite{Lan57} $G =
{\frac {e^2} h} T$.

As expected for mesoscopic systems with discrete energy levels
connected to continuum reservoirs, molecular wires display
resonances in the transmission probability.  Another
potentially important transport phenomenon that has been
predicted in molecular wires is the appearance of
antiresonances.\cite{Kemp96} These are defined to be zeroes of
the transmission and correspond to the incident electrons being
perfectly reflected by the molecule.  In molecular systems this
phenomenon was first recognised in theoretical studies of
electron transfer between donor and acceptor sites of a
molecule.\cite{Ratner90,Cast93,Cheong94} At that time it was
correctly attributed to interference effects between the
different molecular orbitals through which the electron
propagates. However, antiresonances have received less
attention in the context of electrical conduction through
molecular wires connected to metallic contacts and we address
this topic theoretically in the present article.\cite{EmberPRL}

The occurrence of antiresonances has also been reported in
other mesoscopic systems.  They have been found in quantum
waveguides\cite{Sols89,Datta89,Ji92,Tek93,Price93,Porod94,Akis95}
where the transmission displays resonance-antiresonance
structure.  These systems in the form of stub tuners have been
shown to operate as electronic gates.  Antiresonances have
also been proposed to occur in double barrier resonant
tunneling (DBRT) devices.  There they have been explained
using a Fabry-Perot model of the DBRT\cite{Boyk92} and have
also arisen in more sophisticated tight-binding calculations
of the same structures\cite{Boyk91}.  In the above devices it
is the wave nature of the electrons that leads to these
interesting interference effects.  Although the wave nature of
the electrons is also the cause of antiresonances in molecular
wires, it will be seen below that different mechanisms are
responsible for their occurrence in the molecular systems.

We begin by considering a simple model of molecular wires
exhibiting antiresonances that we solve analytically. We then
proceed to investigate the robustness of the analytically
predicted behaviour by studying more realistic models
numerically.

Our analytic model of the molecular wire consists of a molecule
attached to two ideal single-channel 1D leads.  Electrons are
incident from the left lead in only one propagating channel and
scatter through the molecule to the single channel of the right
lead.  The electronic structure of the molecule is described as
a discrete set of molecular orbitals which couple to the single
mode leads. We show that in discussing molecular wire
antiresonances it is important to take into account explicitly
the fact that the atomic orbitals on neighbouring atoms overlap
each other; in some of the systems that we consider
antiresonances are only found if this non-orthogonality is
included fully in the theory. In our analytic calculations the
non-orthogonality is taken into account exactly by defining a
new energy and overlap dependent Hamiltonian in a basis that is
orthogonal and spans a new Hilbert space.\cite{EmberPRL} This
switching of Hilbert spaces greatly simplifies the analytic
solution of the present problem and should have broad
applicability to other transport problems as well, whenever the
mutual non-orthogonality of tight-binding states is important.
It is an alternative to standard orthogonalisation
transformations such as the Wannier or L\"{o}wdin
transformation.  It has the advantages of being much simpler to
implement and much more flexible than the Wannier
transformation since it can be usefully applied to {\em all}
systems described by tight binding models in contrast to the
transformation to Wannier states that is useful mainly in the
theory of crystalline solids. It also differs from the
L\"{o}wdin transformation that is used in quantum chemistry
which defines a new set of orthogonal atomic orbitals in terms
of the original non-orthogonal atomic orbital set.\cite{Lowd50}
We solve the Lippmann-Schwinger equation in this new Hilbert
space to find the transmission probability $T$ of the electrons
to scatter through the molecule.  We derive a simple condition
controlling where the antiresonances occur in the transmission
spectrum.  This condition only depends on the free propagator
for the molecule and the energy- dependent couplings between
the molecular orbitals and the ideal leads.  For a molecule
with $N$ orbitals, the antiresonance condition predicts that
there can be at most $(N-1)+2$ antiresonances.

We then present numerical results for two more general
molecular wire models in order to show how antiresonances
might be observable in real systems.  Our analytic model is
applicable to $\pi$ conjugated systems where the $\pi$
orbitals are independent of the $\sigma$ states that are also
present in realistic systems.  It is able to predict the
energies at which antiresonances occur in our more
sophisticated calculations.  The first of these uses
polyacetylene-like polymers for the two leads which are
attached to a molecule that has a single $\pi$ molecular
orbital.  This molecular wire exhibits a transmission
antiresonance in the occupied $\pi$ band of the leads.

Our second more realistic model is of a molecular wire bridging
a mechanically controlled break junction in a metal wire. In
this case the molecular wire consists of an ``active''
molecular segment connected to the two metal contacts by a pair
of finite $\pi$ conjugated chains. In this calculation we show
how an antiresonance can be generated near the Fermi energy of
the metallic leads.  The differential conductance is calculated
for this system using Landauer theory and the antiresonance is
characterised by a dip in conductance.  We find that for
both of these calculations involving multi-mode leads our
analytic theory of antiresonances has predictive power.

In Sec. II, we describe the method that we use to treat the
non-orthogonality of atomic orbitals and present the solution
to the Lippmann-Schwinger equation for our analytic model.
The antiresonance condition is then derived in Sec. III.  Two
calculations for more realistic systems are presented in
Sec. IV.  We then conclude with Sec. V. 
The Appendix summarises the calculation of the Green's
function for the semi-infinite 1D leads and takes the
non-orthogonality of atomic orbitals into account.

\section{Analytic Theory: Changing Hilbert spaces and solution of the
Lippmann-Schwinger equation}
Most theoretical studies of molecular wires have used tight
binding bases of atomic orbitals for the molecular and lead
Hamiltonians. The atomic orbitals on different atoms are not
orthogonal to each other and, as we will show below, this
non-orthogonality can have important physical consequences for
molecular wire antiresonances.  We treat this lack of
orthogonality exactly in our analytic Lippmann-Schwinger (LS)
theory\cite{LSortho} of antiresonances in a molecule connected
to single-channel leads by solving the problem in a new
Hilbert space spanned by an orthogonal basis where we define a
new energy-dependent Hamiltonian matrix.\cite{EmberPRL} We
begin with a derivation and discussion of this change of
Hilbert space which will be vital to our definition of a LS
equation below.

We start with Schr{\"o}dinger's equation for the eigenvectors
$\{|\Psi\rangle \}$ of a Hamiltonian $H$,
\begin{equation}
H |\Psi\rangle  =  E |\Psi\rangle .
\label{eq:Schrod}
\end{equation}
We wish to solve Eq.(\ref{eq:Schrod}) for $|\Psi\rangle$. We
begin by expressing $|\Psi\rangle$ in a non-orthogonal basis
$\lbrace |n\rangle \rbrace$ of the usual physical Hilbert space
$A$ for the system as $|\Psi\rangle = \sum_n \Psi_n |n\rangle$.
Inserting this expression for $|\Psi\rangle$ into
Eq.(\ref{eq:Schrod}) and applying $\langle m |$ we obtain
\begin{equation}
\sum_n H_{m,n} \Psi_n  =  E \sum_n S_{m,n} \Psi_n
\label{eq:Schrodbasis}
\end{equation}
where we define $H_{m,n}=\langle m |H|n\rangle$ to be the
Hamiltonian matrix and $S_{m,n}=\langle m |n\rangle$ to be the
overlap matrix.
We note that if the basis $\lbrace |n\rangle \rbrace$ is
incomplete then Eq. (\ref{eq:Schrodbasis}) becomes an
approximation that may be justified
variationally.\cite{Lowd50} In either case, we will assume that
Eq.(\ref{eq:Schrodbasis}) provides an adequate description of
the system of interest and our objective will be to solve it
exactly for the coefficients $\Psi_n$.

We will assume in the following that the sums in
Eq.(\ref{eq:Schrodbasis}) (and similar summations in the
remainder of this article) converge absolutely\cite{sums} so
that the order in which the summations are performed is
unimportant. This assumption is justified for the physical
applications that we will be considering where the
non-orthogonal basis states $\lbrace |n\rangle \rbrace$ will
be atomic tight binding orbitals (or molecular orbitals
confined to finite molecular segments) and only a finite
number of these orbitals are considered on any particular
site.  For such basis states $H_{m,n}$ and $S_{m,n}$ decrease
{\em exponentially} as the spatial separation between the
tight binding sites associated with basis states $|m\rangle$
and $|n\rangle$ becomes large. This guarantees the absolute
convergence of the summations in Eq.(\ref{eq:Schrodbasis})
even if the system is infinite in extent and $|\Psi\rangle$ is
a physical scattering state that extends throughout the
system.

The absolute convergence of the series in
Eq.(\ref{eq:Schrodbasis}) enables us to rewrite
Eq.(\ref{eq:Schrodbasis}) as
\begin{equation}
\sum_n H^E_{m,n} \Psi_n  =  E \Psi_m
\label{eq:Schrodbasis'}
\end{equation}
where
\begin{equation}
H^E_{m,n} = H_{m,n} - E(S_{m,n}-\delta_{m,n}).
\label{eq:H'}
\end{equation}
We are concerned with open systems that have a continuous
spectrum of energy eigenvalues. Thus our objective is to find
the coefficients $\Psi_n$ that satisfy
Eq. (\ref{eq:Schrodbasis'}) for every value of $E$ belonging
to the continuum of energy eigenvalues of the Hamiltonian
$H$. To do this we find it convenient to consider the related
matrix eigenproblem
\begin{equation}
\sum_n H^E_{m,n} \Psi'_n  =  E' \Psi'_m
\label{eq:SchrodbasisE'}
\end{equation}
where $E'$ is any eigenvalue of the matrix $H^E_{m,n}$ and the
set of coefficients $\lbrace \Psi'_m \rbrace$ form the
corresponding eigenvector. The solution $\lbrace \Psi_m
\rbrace$ of Eq. (\ref{eq:Schrodbasis'}) that we seek is then
identical to an eigenvector $\lbrace \Psi'_m \rbrace$ of the
matrix $H^E_{m,n}$ for which the eigenvalue $E'$ in
Eq. (\ref{eq:SchrodbasisE'}) is equal to $E$.

We now re-interpret Eq. (\ref{eq:SchrodbasisE'}) as the matrix
form of a new Schr{\" o}dinger equation
\begin{equation}
H^E |\Psi\rangle'  =  E' |\Psi\rangle' .
\label{eq:Schrod'}
\end{equation}
involving a new
Hamiltonian operator $H^E$ and its eigenvectors
$|\Psi\rangle'$ in a new Hilbert space
$A'$. We  construct this new Hilbert space as follows: We first
form the vector space $V'$ that is spanned by an {\em
orthonormal} basis (that we denote $\{ |n\rangle' \}$) of the
matrix
$H^E_{m,n}$.\cite{basis} [Note that the basis vectors
$|n\rangle'$ defined in this way are abstract mathematical
entities which should not be confused with the (non-orthogonal)
physical state vectors
 $|n\rangle$ of the original (physical) Hilbert space $A$ or
with any other vectors in that Hilbert space.] For the systems
of interest in this work $V'$ is infinite-dimensional and we
define  the Hilbert space
$A'$ to be the completion of $V'$ with respect to the norm
topology.\cite{hilbert}  Eq. (\ref{eq:SchrodbasisE'}) will be
the matrix form of Eq. (\ref{eq:Schrod'}) in Hilbert space $A'$
as desired provided that the new Hamiltonian operator
$H^E$ is chosen so that its matrix elements satisfy
$\langle m|' H^E |n\rangle' = H^E_{m,n}$ and
$|\Psi\rangle' = \sum_{n} \Psi'_{n} |n\rangle'$.
It follows from Eq.(\ref{eq:H'}) that the operator $H^E$ is
Hermitian in $A'$ because $E$ is real and $H_{m,n}\,$,
$S_{m,n}$ and $\delta_{m,n}$ are Hermitian matrices.

Thus we have transformed a problem that was formulated in
terms of a non-orthogonal basis into an equivalent one in an
orthogonal basis of a {\em different} Hilbert
space. Essentially what we have done is to create a new
problem (which may be easier to solve) from our old one.  This
is quite different from other orthogonalisation schemes (such
as that of Gramm-Schmidt or L\"{o}wdin\cite{Lowd50}). In those
schemes the original non-orthogonal basis of the Hilbert space
is orthogonalised by transforming it into a new orthogonal
basis of the same space.  Our method has no such
orthogonalisation procedure: instead we assume our new
operators and eigenvectors to be expressed in terms of an
orthogonal basis of a new space and define them so that the
matrix eigenvalue problem, Eq. (\ref{eq:SchrodbasisE'})
follows. This re-definition creates an energy dependent
Hamiltonian whose energy dependence will be important in our
discussion of antiresonances below. 

It should be noted that only the eigenvectors of $H^E$ that
have the eigenvalue $E$ have the same coefficients $\Psi_n$ as
eigenvectors of the true Hamiltonian $H$. The other
eigenvectors of $H^E$ do not correspond to any eigenstate of
the physical Hamiltonian $H$, but they never the less play an
important role when calculating the Green's function
corresponding to $H^E$ which appears in the Lippmann-Schwinger
equation below.

Since no assumptions (other than the absolute convergence of
the summations in Eq.(\ref{eq:Schrodbasis})) have been made
about the nature of the system being considered, this method
of orthogonalisation by switching to a new Hilbert space is
extremely general. If the basis states $\lbrace |n\rangle
\rbrace$ are tight binding atomic orbitals then the present
transformation (unlike the transformation to Wannier
functions) can be used irrespective of the types of atoms
involved or their locations in space. Furthermore, our
transformation has the additional flexibility that the
non-orthogonal basis states need not all be of the same
generic type. For example, some of them may be atomic orbitals
and others molecular orbitals on some cluster(s) of atoms that
form a part of the physical system. This flexibility will be
exploited below.  We now proceed to outline the application to
antiresonances in molecular wires.

\begin{figure}[ht]
\includegraphics[bb = 0 0 640 800, width = 
0.75\textwidth,clip]{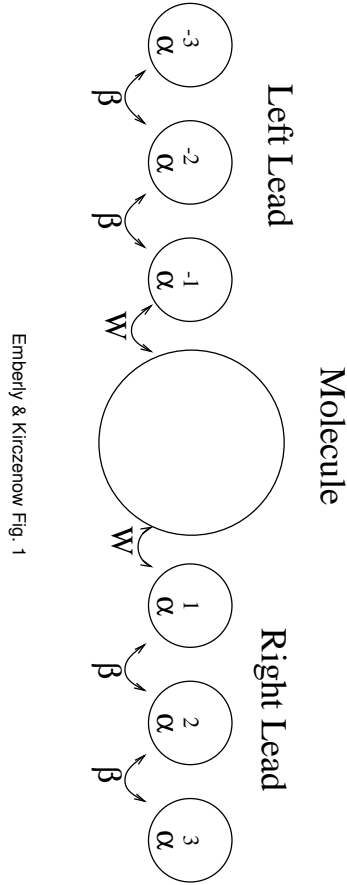}
\begin{center}
\caption{A schematic diagram for the idealized model of a
molecular wire, consisting of left and right single channel
leads and the molecule. These three systems are described by
the Hamiltonian $H_0$, with $\langle n|H_0|n\rangle = \alpha$
and $\langle n|H_0|m\rangle = \beta$ for $n,m$ on the left or
right leads with $m=n\pm1$. On the molecule $\langle
\phi_j|H_0|\phi_k\rangle = \epsilon_j \delta_{j,k}$.  The
molecular orbitals are coupled to the adjacent lead sites by
$W$, with $\langle -1|W|\phi_j\rangle = W_{-1,j}$ etc..  }
\label{fig1}
\end{center}
\end{figure}

Our analytic theory for electron transport in molecular wires
is based on an idealised model consisting of a molecule
connected to two identical single-channel ideal leads which
are represented by 1D chains of atoms (shown in
Fig. \ref{fig1}).  We solve for the scattering wavefunction
$|\Psi\rangle$ which describes an electron incident from the
left lead with energy $E$ and having a probability $T(E)$ to
scatter through the molecule to the right lead.  The system
satisfies $H |\Psi\rangle = E |\Psi\rangle$, where $H = H_0 +
W$.  $H_0$ is the Hamiltonian for the three isolated systems
consisting of the two leads and the molecule and $W$ couples
the lead sites adjacent to the molecule with the sites on the
molecule.

We now introduce a non-orthogonal basis consisting of atomic
orbitals $\{|n\rangle\}$ with $n=-\infty,\ldots,-1$ on the
left lead and $n=1,\ldots,\infty$ on the right lead and a set
of molecular orbitals (MO's) $\{|\phi_j\rangle\}$ for the
molecule.  In this basis the wavefunction has the form,
\begin{equation}
|\Psi\rangle = \sum_{n=-\infty}^{-1} \Psi_{n} |n\rangle +
\sum_{n=1}^{\infty} \Psi_{n} |n\rangle +  \sum_j c_j
|\phi_j\rangle
\end{equation}
The transmission probability $T$ is found from $|\Psi\rangle$
by utilising the boundary conditions that the wavefunction
satisfies.  On the left lead the wavefunction consists of a
rightward propagating Bloch wave along with a reflected
leftward propagating Bloch wave.  This can be written as
$|\Psi_L\rangle = \sum_{n=-\infty}^{-1} (\exp(iny) + r
\exp(-iny))|n\rangle$, where $r$ is the reflection
coefficient.  The right lead is identical to the left lead and
on it the wavefunction has the form of a transmitted Bloch
wave $|\Psi_R\rangle = \sum_{n=1}^{\infty} t \exp(iyn)|n\rangle$
where $t$ is the transmission coefficient.  Thus the
transmission probability that enters the Landauer electrical
conductance of the wire is given by $T = |t|^2 =
|\Psi_1|^2$. The electron's velocity is the same on both leads
and so the ratio of velocities that normally appears in the
formula for $T$ is equal to unity.

In the non-orthogonal basis solving for $\Psi_1$ analytically
is difficult thus we change to the new Hilbert space where the
solution is more straightforward.  The transmission
probability $T$ is unaffected since the coefficient $\Psi'_1$
remains the same for fixed $E$. The new Hamiltonian operator
$H^E$ and its eigenvectors $\{|\Psi\rangle ' \}$ are now
assumed to be expressed in an orthonormal basis
$\{|n\rangle',|\phi_j\rangle\}$ with the new Hamiltonian
matrix elements defined in terms of the matrix elements of the
initial Hamiltonian $H$ and the overlap matrix $S$ via
Eq. (\ref{eq:H'}). Thus if there is any non-orthogonality in
the original basis sets of the three isolated systems then
$H_0$ becomes $H^E_0$.  The non-orthogonality between the
orbitals on the molecule and the leads changes $W$ to $W^E$.

We evaluate $\Psi'_1$ by solving a Lippmann-Schwinger (LS)
equation.  This equation is defined only after the
transformation and is given by,
\begin{equation}
|\Psi\rangle' = |\Phi_{o}\rangle' + G_{o}(E) W^E |\Psi\rangle'
 .
\label{eq:LS}
\end{equation}
Here $G_{o}(E) = (E - H_0^E)^{-1}$ is the Green's function for
the decoupled system of left, right leads and the molecule.
The electron is initially in the eigenstate $|\Phi_{o}\rangle'$
of the left lead propagating with an energy $E$.  It is
confined to the left lead and is written as
\begin{equation}
|\Phi_{o}\rangle' = \sum_{n=-\infty}^{-1} (\Phi'_{o})_{n}
 |n\rangle' .
\end{equation}

It should be emphasised that the LS equation Eq. (\ref{eq:LS})
is only valid after the change to the new Hilbert
space where the basis is orthogonal.  This is because it is
now possible to distinctly separate the states on the leads
from those on the molecule.  The original non-orthogonal basis
did not allow for this clear distinction and contradictions
arise if the analogs of the entities in Eq. (\ref{eq:LS}) are
constructed using this basis.

The free propagator, $G_{o}$, can be expressed in terms of the
energy eigenstates of $H_{0}^E$ of the isolated leads and of
the molecule. It will be written as a sum of three separate
free propagators for the left and right leads and the
molecule, $G_{o}^{L}$, $G_{o}^{R}$, and $G_{o}^{M}$
respectively.  The left and right leads have been assumed to
be identical so their free propagators will be the same.  For
the leads with energy eigenstates $\{|\Phi_{o}(y)\rangle'\}$
having energy $\epsilon(y)$
\begin{equation}
G_{o}^{R} = \sum_y \frac{|\Phi_{o}(y)\rangle'\langle
\Phi_{o}(y)|'}{E-\epsilon(y) }
\end{equation}
Here $y$ is the wave number in units of the inverse lattice
parameter. Expressing the eigenstates in terms of the basis,
$\{|n\rangle'\}$, the free propagator on the leads has the
form,
\begin{equation}
G_{o}^{R} = \sum_{n=1}^{\infty} \sum_{m=1}^{\infty}
(G_{o}^{R})_{n,m}|n\rangle' \langle m|'
\end{equation}
The matrix elements, $(G_{o}^{R})_{n,m}$ are evaluated
analytically in the Appendix.

For the molecule the free propagator expressed in terms of its
MO's is
\begin{equation}
G_{o}^{M} = \sum_j \frac{|\phi_j\rangle \langle \phi_j
|}{E-\epsilon_j} = \sum_j (G_{o}^{M})_j |\phi_j\rangle \langle
\phi_j |
\end{equation}

Using these expressions for the wavefunctions and the free
propagator the LS equation becomes (the $'$ have been dropped)
\begin{eqnarray*}
\sum_{n=-\infty}^{-1} \Psi_n |n\rangle + \sum_{n=1}^{\infty}
\Psi_n |n\rangle + \sum_j c_j |\phi_j\rangle =
\sum_{n=-\infty}^{-1} (\Phi_{o})_n |n\rangle + \\
(\sum_{n,m=-\infty}^{-1} (G_{o}^{L})_{n,m} |n\rangle \langle
m| + \sum_{n,m=1}^{\infty} (G_{o}^{R})_{n,m} |n\rangle \langle
m| + \sum_j (G_{o}^{M})_j |\phi_j\rangle \langle \phi_j| )
\,W^E\,\times
\\
(\sum_{n=-\infty}^{-1} \Psi_n |n\rangle + \sum_{n=1}^{\infty}
\Psi_n |n\rangle + \sum_j c_j |\phi_j\rangle)
\end{eqnarray*}
We now apply the bras $\langle -1|$, $\langle \phi_j|$, and
$\langle 1|$ to the above equation, making use of their formal
mutual orthogonality.  This gives the following set of
simultaneous linear equations
\begin{eqnarray}
\Psi_{-1} &=& (\Phi_{o})_{-1} + (G_{o}^{L})_{-1,-1} \sum_j
\langle -1 | W^E |\phi_j \rangle c_j \\
c_j &=& (G_{o}^{M})_j(\langle \phi_j |W^E|-1\rangle \Psi_{-1} +
\langle \phi_j |W^E|1\rangle \Psi_{1}) \\
\Psi_{1} &=& (G_{o}^{R})_{1,1} \sum_j \langle 1|W^E|\phi_j\rangle
c_j
\end{eqnarray}
where
\begin{equation}
\langle 1|W^E| \phi_j\rangle = W^E_{1,j} = W_{1,j} - E S_{1,j}
\end{equation}
which is the interaction matrix element between the the lead
orbital adjacent to the molecule and the $j^{th}$ MO.  Notice
it includes the overlap between the $j^th$ MO and the first
lead orbital.

These equations can be solved for the unknowns, $\Psi_{-1}$,
$\Psi_{1}$, and the $c_j$, yielding
\begin{eqnarray}
\Psi_{1} &=& \frac{A (\Phi_o)_{-1})}{[(1-B)(1-C)-AD]}
\label{eq:psi1} \\
\Psi_{-1} &=& \frac{(1-B)(\Phi_o)_{-1}}{[(1-B)(1-C)-AD]} \\
c_j &=& (G_{o}^{M})_j \left( \frac{W^E_{j,1}A +
W^E_{j,-1}(1-B)}{[(1-B)(1-C) - AD]} \right) (\Phi_{o})_{-1}
\end{eqnarray}
where, making use of the symmetry between $G_{o}^{L}$
and $G_{o}^{R}$,
\begin{eqnarray*}
A &=& (G_{o}^{R})_{1,1} \sum_j W^E_{1,j} (G_{o}^{M})_j W^E_{j,-1} \\
B &=& (G_{o}^{R})_{1,1} \sum_j (W^E_{1,j})^2 (G_{o}^{M})_j \\
C &=& (G_{o}^{R})_{1,1} \sum_j (W^E_{-1,j})^2 (G_{o}^{M})_j \\
D &=& (G_{o}^{R})_{1,1} \sum_j W^E_{-1,j} (G_{o}^{M})_j W^E_{j,1}
\end{eqnarray*}
The transmission probability $T(E)$ is given by $|\Psi_1|^2$.

\section{Antiresonance Condition and Mechanisms}
An antiresonance is defined to be a zero of the electron
transmission probability.  Since $T(E)$ is given by the
squared modulus of $\Psi_1$, the zeroes occur
where $\Psi_1$ is zero.  From Eq. (\ref{eq:psi1}) this happens
when $A=0$.  Thus the antiresonance condition is
\begin{equation}
(G_{o}^{R})_{1,1} \sum_j W^E_{1,j}(G_{o}^{M})_j W^E_{j,-1} = 0
\label{eq:anti:gen}
\end{equation}
or
\begin{equation}
\sum_j \frac{(W_{1,j} - E S_{1,j})(W_{j,-1} - E
S_{j,-1})}{E - \epsilon_j} = 0
\label{eq:anti}
\end{equation}
where the sum over $j$ includes just the MO's.

The antiresonance conditions (\ref{eq:anti:gen}) and
(\ref{eq:anti}) that we have derived allows us to identify two
distinct mechanisms that can give rise to antiresonances in
molecular wire transport.

In the first of these mechanisms, antiresonances arise due to
an interference between the different MO's of the molecule.
This is seen directly from Eq. (\ref{eq:anti}): An electron
incident from the left lead, hops from the lead site adjacent
to the molecule onto each of the molecular orbitals with a
weight $W^E_{j,-1}$.  It then propagates through each of the
different orbitals $j$.  These processes interfere with each
other as the electron propagates through the molecule and
proceeds to hop onto the first lead site on the right lead with
a weight $W^E_{1,j} \, .$

What is particularly interesting about this molecular mechanism
and differentiates it qualitatively from a standard multi-beam
interference problem encountered in optics via a diffraction
grating, is that the antiresonances arise from {\em
interference between molecular states that differ in energy}.
It is also not possible to make an analogy between this effect
and Fano resonances\cite{Fano61}, which are a good analog of
the antiresonances in electron waveguides of the stub-tuner
type \cite{Porod94}.  For those waveguides, the antiresonances
arise from interference between the direct transmission of a
continuum of electron modes (which exists in the semiconductor
quantum wire) and transmission via discrete modes that reside
within the resonator.  In our model, transmitted electrons {\em
must} pass through the molecule so that there is no {\em
direct} transmission of continuum modes from the left to the
right lead and the Fano mechanism does not apply.  The
molecular wire antiresonances are also not analogous to those
found in the Fabry-Perot model of double barrier resonant
tunneling\cite{Boyk92}.  In that model, electrons couple to
different modes within the well which interfere upon exiting
the well.  But these modes are all at the same energy.  As was
mentioned above, the interfering molecular states are at
different energies.

This antiresonance mechanism is qualitatively similar to that
which was found in previous work on electron transfer between
donors and acceptors in molecules,\cite{Ratner90} however the
antiresonance conditions (\ref{eq:anti:gen}) and
(\ref{eq:anti}) that we have derived differ from those that
were obtained earlier partly because the non-orthogonality of
the atomic orbitals has been included in our theory. In
particular, for a molecule with N distinct energy levels, the
resonance condition (\ref{eq:anti}) gives rise to a polynomial
equation of degree $m = (N-1) + 2$, so there can be at most $m$
antiresonances for this model. Neglecting the overlaps
$S_{1,j}$ and $S_{j,-1}$ between atomic orbitals on different
atoms leads to a polynomial equation of a lower order and can
yield qualitatively different predictions, as will be made
clear below.

The second antiresonance mechanism that we identify on the
basis of the conditions (\ref{eq:anti:gen}) and (\ref{eq:anti})
has no analog in previous work and is at first sight quite
surprising since it is due entirely to the non-orthogonality of
atomic orbitals on different atoms. It occurs when only a
single molecular orbital $a$ couples appreciably to the leads,
which should happen in some real systems for reasons of
symmetry, as is discussed in the next section.  In such cases
Eq. (\ref{eq:anti}) becomes $(W_{a,-1} - E
S_{a,-1})(W_{1,a} - E S_{1,a}) = 0$. As many as two
antiresonances are possible in this case.  They arise because
the energy dependent coupling is equal to zero between the
leads and the molecule at the energies $E$ for which the energy
dependent hopping parameter $(W_{a,-1} - E S_{a,-1})$ or
$(W_{1,a} - E S_{1,a})$ vanishes.  This cancellation arises
because of the non-orthogonality between the atomic orbitals of
the molecule and those of the leads.

We explore some specific molecular systems that should exhibit
each of the above antiresonance mechanisms in the following
section.

The present theory is readily extended to include second and
more distant neighbour interactions and overlaps. These will act
as a perturbations to the antiresonance values. If the
interactions are sufficiently long ranged so as to couple the
two leads to each other directly in addition to coupling them
indirectly via the molecule, other antiresonance mechanisms,
including Fano-like effects become possible. However detailed
consideration of these is beyond the scope of this paper.

\section{Multichannel Leads and Metallic Contacts}
The above model has yielded an equation which predicts
energies at which antiresonances should occur in molecular
wire systems.  It was based on a highly idealised model which
assumed that there was only a single propagating electronic
mode in the leads.  This single mode was only allowed to
interact with the molecular orbitals on the molecule.  Real
leads, whether organic or inorganic, will certainly not be as
simple.  However the two calculations presented below show
that the key predictions made by this simple model should apply
quantitatively to some more complex systems as well.

A good approximation to a 1D lead with only one orbital per
site is conjugated trans-polyacetylene.  The $\pi$ backbone of
this polymer is orthogonal to the $\sigma$ orbitals in the
plane.  Second nearest neighbour $\pi$ interactions between
carbon atoms are also small compared to nearest neighbour
interactions.  The conjugation of the polymer creates a band
gap in the $\pi$ energy band of this system.  If one inserts a
molecule whose spectrum also consists of $\pi$ and $\sigma$
molecular orbitals into the backbone of this structure in a
suitably symmetric way, only the $\pi$ band of the polymer
will interact with the $\pi$ states of the molecule.  It is
important however that the inserted molecule be long enough so
that the $\pi$ orbitals on the left polyacetylene lead can not
overlap with the $\pi$ orbitals on the right lead.  Otherwise
electrons could hop directly from the left lead to the right
lead without passing through the inserted molecule.  Thus our
simple model is applicable to a system consisting of $\pi$
conjugated leads attached to a molecular wire with $\pi$
states if the wire is long enough that there is no direct
interaction between the leads and the geometry is such that
$\sigma$ - $\pi$ hybridization between the leads and molecular
wire is forbidden. The antiresonance condition
Eq. (\ref{eq:anti}) should be able to predict the
antiresonances of these more complicated systems if the
energies $\epsilon_j$ of the $\pi$ molecular orbitals are
specified along with their interaction energies $W_{j,1}$ and
$W_{-1,j}$ with the $\pi$ mode of the leads and the
corresponding orbital overlaps $S_{j,1}$ and $S_{-1,j}$.

We now proceed to calculate the electron transmission
probability for two such model systems. Our calculations are
based on a numerical method which determines the transmission
probability of a molecular wire coupled to multichannel
tight-binding leads.\cite{Ember98,Lead} The multichannel leads
are constructed out of multi-atom unit cells which are then
attached to the molecular wire.  The non-orthogonality within
the leads and the molecule and also between the leads and the
molecule is treated with the use of our transformation.  The
calculation proceeds by evaluating the band structure of the
left and right leads from which it is then possible to
determine the propagating electron modes (Bloch states)
$\{|\Phi^+_j\rangle,\Phi^-_j\}$ at a given energy $E$.  In
these multichannel calculations the wavefunction for an
electron incident in the $i^{th}$ mode has the following
boundary conditions.  In the left lead, $|\Psi_L^i\rangle =
|\Phi^+_i\rangle + \sum_j r_{j,i} |\Phi^-_j\rangle$. On the
molecule the wavefunction is a linear combination of the atomic
orbitals on the molecule.  On the right lead the wavefunction
is $|\Psi_R^i\rangle = \sum_j t_j,i |\Phi^+_j\rangle$. With the
above form for the wavefunction we then solve Schr\"{o}dinger's
equation $H|\Psi^i\rangle = E |\Psi^i\rangle$ for the molecular
wire system to find the transmission amplitudes $t_{j,i}$ which
connect the modes $i$ in the left lead to those in the right
lead, $j$.  The transmission probability is then found using
\begin{equation}
T(E) = \sum_i \sum_j \left | \frac{v_i}{v_j} \right |
|t_{i,j}|^2
\label{eq:multiT}
\end{equation}
where the sum over $j$ is over the rightward propagating modes
in the left lead and the sum over $i$ is over rightward
propagating modes in the right lead.  The velocity ratio now
appears since the velocities of modes in the left and right
leads may be different.  We now show that provided the
model system meets the assumptions of the analytical model
presented above, this more sophisticated numerical method
yields results consistent with the analytical predictions.

\begin{figure}[ht]
\includegraphics[bb = 0 0 640 800, width = 
0.75\textwidth,clip]{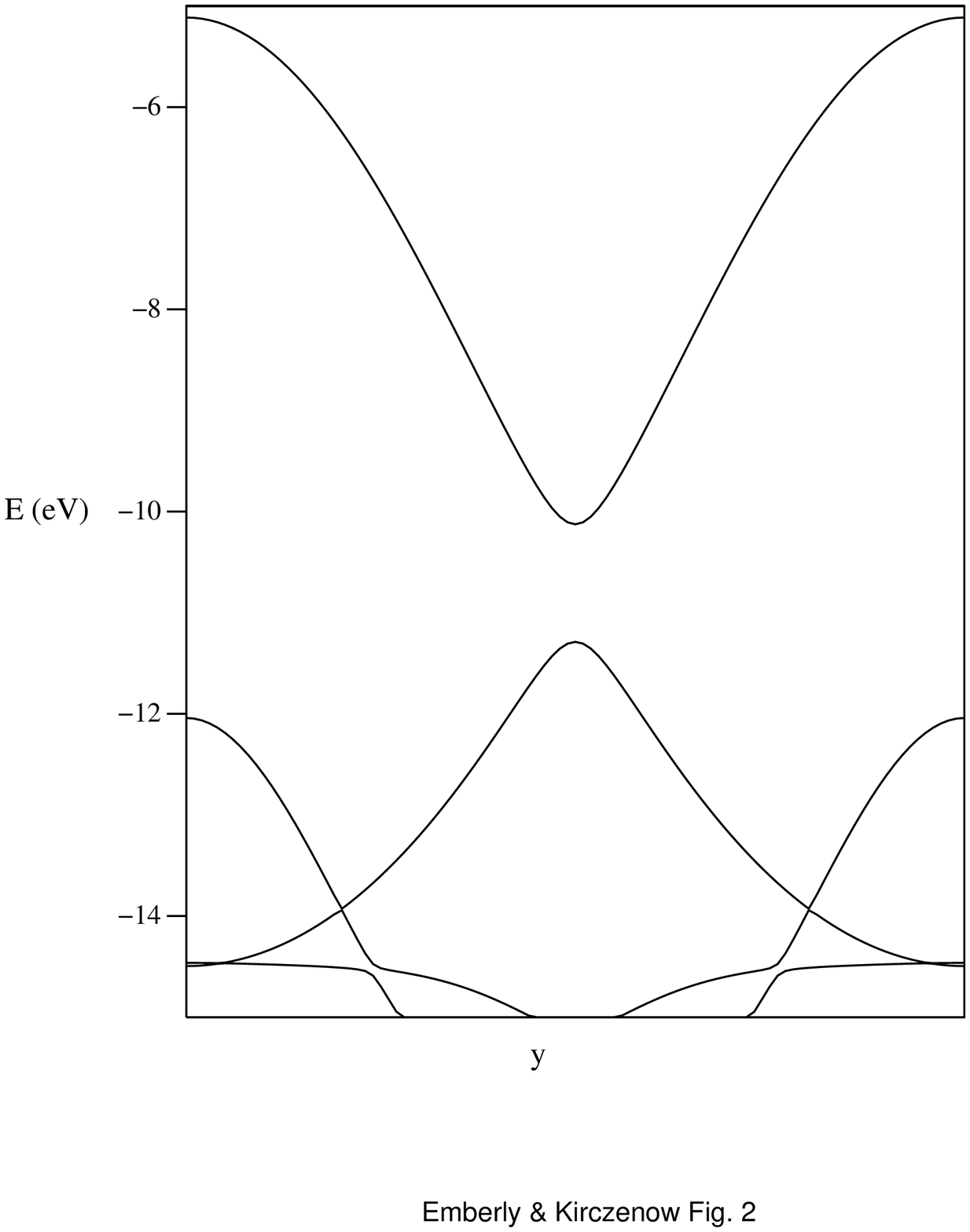}
\begin{center}
\caption{Band structure for conjugated polyacetylene calculated
using extended H\"{u}ckel.  The $\pi$ band extends from -14.5
eV to -5 eV and has a band gap starting at -11.2 eV. }
\label{fig2}
\end{center}
\end{figure}

\begin{figure}ht]
\includegraphics[bb = 0 0 640 800, width = 
0.75\textwidth,clip]{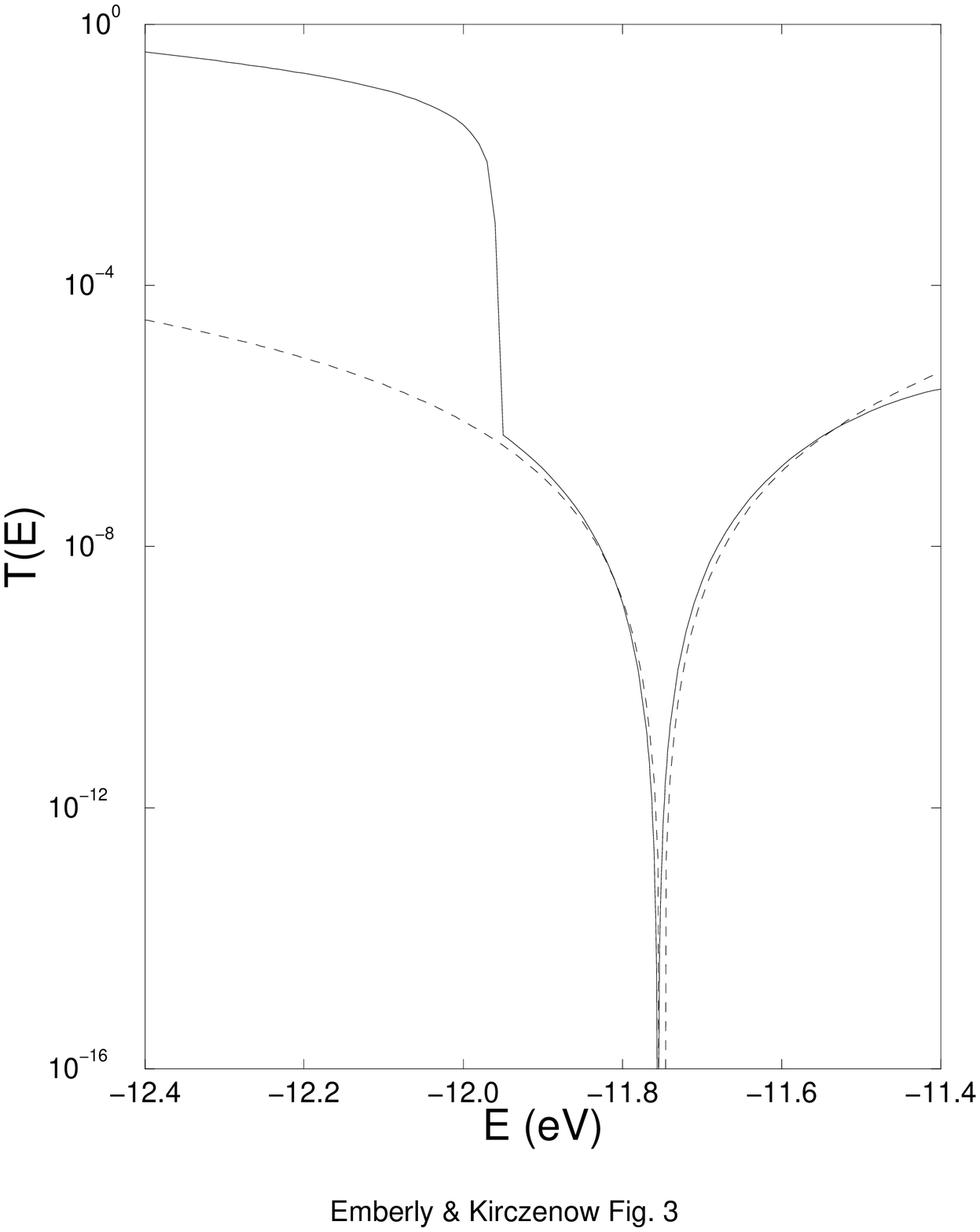}
\begin{center}
\caption{(Solid line) transmission plot for a molecule with a
single $\pi$ state connected to polyacetylene leads.  (Dashed
line) transmission plot calculated using analytic theory for
similar system.  }
\label{fig3}
\end{center}
\end{figure}

The first calculation utilises the trans-polyactylene polymer
to model the left and right ideal leads. We have calculated the
band structure for these leads and this is shown in
Fig. \ref{fig2}.  The unit cell for the polyacetylene lead was
taken to consist of two CH groups and the group spacing was
taken from Su {\it{et al}}\cite{Su79}.  The $\pi$ band extends
from around -14.5 eV to -5 eV and has a band gap of around -1.4
eV starting at -11.2 eV.  The upper $\pi$ band is unoccupied.
The other energy bands are $\sigma$ modes. In reality, it is
well known from electron transport studies of
trans-polyacetylene that soliton and polaron formation is
favoured when charge is injected into the
chain.\cite{Su79,Su80} We do not include such effects in the
present calculation; here the infinite polyacetylene chains are
taken to be static periodic structures, they represent ideal
quasi-one dimensional leads.  For this system we consider a
molecular wire with three $\sigma$ states but with just a
single $\pi$ level to which the leads couple.  The model
parameters for the molecular wire's coupling of its $\pi$ state
to the $\pi$ band of the leads are chosen so that our analytic
antiresonance condition (\ref{eq:anti}) predicts an
antiresonance in the occupied $\pi$ energy band and at an
energy where only a $\pi$ mode propagates in the left lead.  If
there were a transmitted $\sigma$ mode present at the energy of
the antiresonance as well its transmission would be
superimposed on the $\pi$ transmission which would possibly
obscure the antiresonance.  A single $\pi$ mode propagates
through this system between the energies of -12 eV and -11.2
eV.  For our molecule with a single $\pi$ state the
antiresonance arising out of the interaction with the left lead
is predicted by (\ref{eq:anti}) to occur at an energy of $E =
W_{\pi,-1}/S_{\pi,-1}$. If we choose the overlap to be 0.3 and
an interaction energy of -3.525 eV between the $\pi$ state of
the molecule and the $\pi$ orbital on the nearest carbon atom
on the lead, the antiresonance is predicted to occur at -11.75
eV.  Since the interaction energy is related to the overlap the
two are not completely arbitrary, however the interaction
energy also depends on the energy of the molecular orbital
which in principle could be chosen to yield the above coupling
energy.  The numerically calculated electron transmission
probability for this system is shown by the solid line in
Fig. \ref{fig3}.  The antiresonance is clearly seen at -11.75
eV.  There is also a sharp drop off in transmission at -12 eV
where there are no longer any $\sigma$ modes incident from the
left lead.  Also shown (the dashed line in Fig. \ref{fig3}) is
the result of the analytic calculation using the (LS) equation
above.  This was done for a single mode lead with its energy
band spanning the width of the polyacetylene $\pi$ band.  The
couplings and overlaps were chosen to be the same as those used
in the multi-mode case.  The agreement between the two
calculations is quite good in the vicinity of the
antiresonance.

The above calculation was for infinite polyactylene leads which
are known to be insulators.  Thus performing a conductance
experiment on such a system would be impossible.  Recent
experiments on molecular wires have used metallic nanocontacts
connected to the molecule\cite{Reed97}.  The system that we
base our next calculation on is such a mechanically controlled
break junction (MCBJ) which is bridged by a single molecule.
The metallic contacts will be taken to be gold.  We consider
the molecular wire to consist of left and right $\pi$
conjugated chain molecules attached to what we will call the
``active'' molecule.  The purpose of these conjugated chains is
to act as a filter to the many modes that will be incident from
the metallic leads.  For appropriate energies they will
restrict the propagating electron mode to be only $\pi$ like.
The electronic structure of the active molecule will be assumed
to consist of $\pi$ and $\sigma$ like molecular orbitals.  The
$\pi$ backbone of the chain molecules will only interact with
the $\pi$ orbitals of the active molecule and so our
antiresonance condition should still be applicable in this
model.

\begin{figure}[ht]
\includegraphics[bb = 0 0 640 800, width = 
0.75\textwidth,clip]{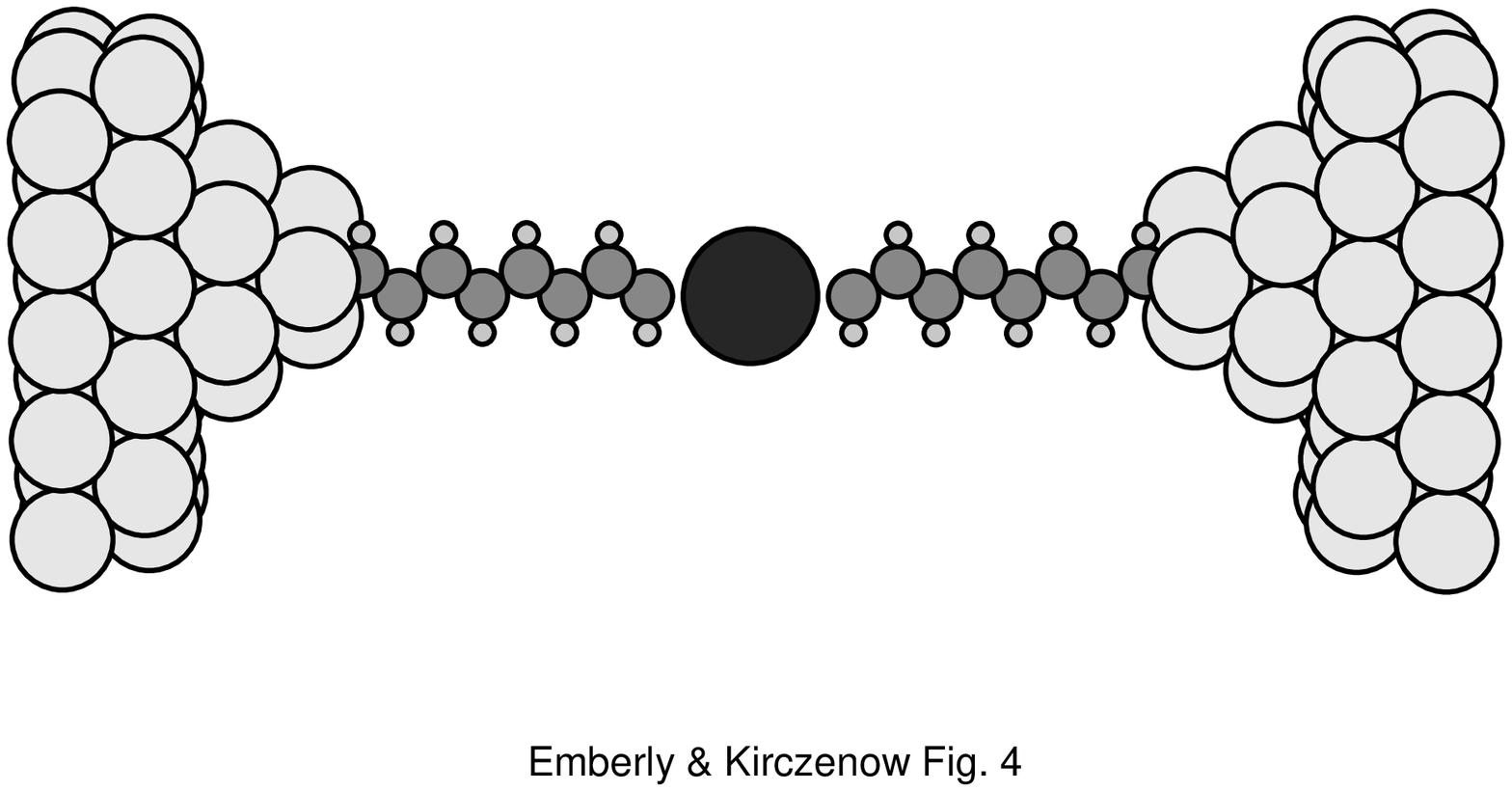}
\begin{center}
\caption{Atomistic diagram for the MCBJ and molecular wire
system.  The first unit cells of the left and right (111) leads
are shown as the last two layers of gold atoms on either side.
Also shown are the (CH)$_8$ chain molecules and ``active''
molecule.  These are attached to two clusters of 10 gold atoms
that form the tip.  The perpendicular distance between the the
last C atoms on the chains and the triangle of gold atoms is
1.6 Angstroms.}
\label{fig4}
\end{center}
\end{figure}

The multi-channel gold leads for our calculation are created
using a unit cell composed of two layers of gold atoms in the
(111) direction.  Both layers have 20 gold atoms.  Since the
Fermi energy for gold resides in the 6s band we only use 6s
orbitals on our gold atoms.  The chain molecules consist of
eight CH groups each.  An atomic diagram of our system is shown
in Fig. \ref{fig4}.  The chain molecules are bonded to clusters
consisting of 10 gold atoms that form the tips of the leads.
The carbon atom nearest to the gold tip binds over the triangle
of gold atoms with a perpendicular distance of 1.6
Angstroms. This larger molecular segment consisting of the gold
tips, chain and active molecules is then attached to the left
and right unit cells of the multichannel gold leads. The chains
are now finite polyacetylene and so they now have discrete
energy states rather than bands, the molecular states on these
chains are $\pi$ like for energies in the polyacetylene $\pi$
band.  This gives rise to the filtering process mentioned
above.  The active molecule is chosen to have two $\sigma$
states and two $\pi$ states.  Unlike the infinite leads in the
preceding calculation the finite chain molecules will conduct.
Their $\pi$ like orbitals will only couple to the two $\pi$
states of the active molecule.

The Fermi energy for our gold leads is around -10 eV which
lies within the $\pi$ band.  Thus we would like an
antiresonance to occur somewhere near this energy.  Again, for
this model system, we make some ad-hoc choices for our
parameters.  The two active molecular $\pi$ states are chosen
to have energies $\epsilon_a = -14.0$ eV and $\epsilon_b =
-11.0$ eV.  The interactions between these states and the
$\pi$ orbitals on the carbon atoms directly adjacent to the
active molecule are $W_{a,-1} = W_{1,a} = -3.0$ eV, and
$W_{b,-1} = -W_{1,b} = 1.25$ eV.  The overlaps are $S_{a,-1} =
S_{1,a} = 0.14$ and $S_{b,-1} = -S_{1,b} = -0.2$.  Solving the
cubic equation to which Eq. (\ref{eq:anti}) reduces in this
case, yields three real energies, one of them in the desired
energy range predicting an antiresonance at -10.08 eV. We now
proceed to calculate the electronic transmission through this
system. 

\begin{figure}[ht]
\includegraphics[bb = 0 0 640 800, width = 
0.75\textwidth,clip]{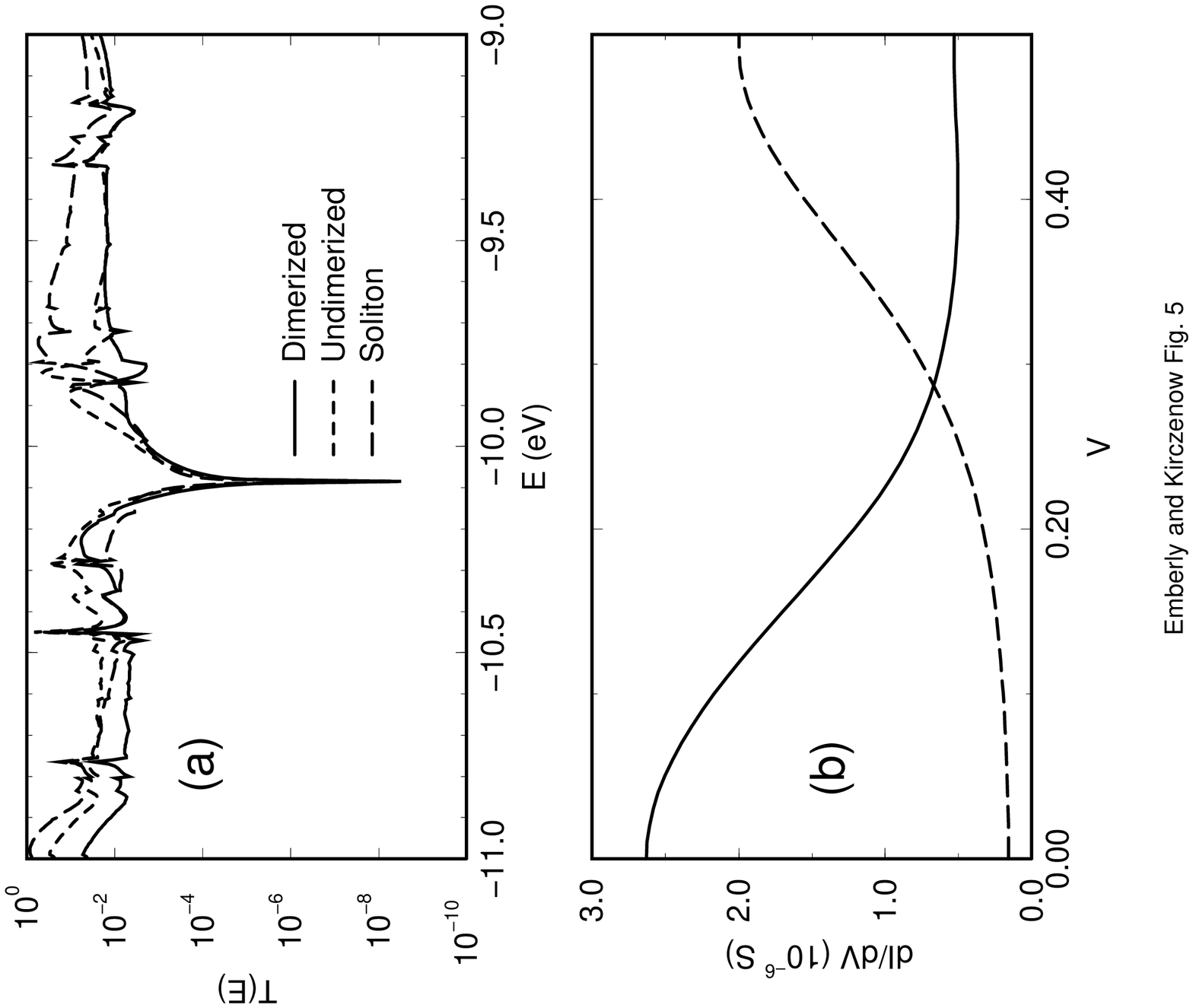}
\begin{center}
\caption{(a) The calculated transmission for a two state active
molecule attached to CH chains and gold (111) contacts.  (b)
The calculated differential conductance assuming a Fermi energy
of (solid line) -10.2 eV and (dashed line) -10.0 eV and a temperature of 
$T=293$ K.} \label{fig5}
\end{center}
\end{figure}

Recent work has studied electron transport through finite
conjugated chains attached to metallic leads with the inclusion
of inelastic degrees of freedom.\cite{Ness99} In that study it
was shown that electron injection onto the chains induces a
small polaron defect. However, for chains of small enough
length (10 CH groups or less) it was found that this polaron
defect did not have an appreciable effect on the electron
transport compared to the static case. To see how atomic
positional disorder on the chains affects the antiresonance, we
have numerically calculated the transmission probability for
several different static atomic configurations of the finite
poly-acetylene chains and these are shown in
Fig. \ref{fig5}a. The solid curve corresponds to dimerized
trans-polyacetylene.  The short-dashed curve corresponds to an
undimerized chain.  The long-dashed curve corresponds to chains
with a static soliton.  In all three curves the antiresonance
is present, although the magnitude varies in the regions
outside of the antiresonance.  Although dynamic effects of
polaron or soliton formation have not been included in the
present study, we have shown that the existence of the
antiresonance is not affected by the disorder on the conjugated
chains or by whether or not they dimerize.  Based on the
results presented in Ref. \cite{Ness99} for short conjugated
chains, we expect that the antiresonance would still survive
even with the inclusion of dynamical effects where the
couplings and overlaps may vary.

In experiments on molecular wires the electric current $I$
through the molecule is measured as a function of voltage
$V$. The differential conductance is then determined by taking
the derivative $G = dI/dV$. We calculate this differential
conductance by using a generalisation of the Landauer formula
which relates the electric current to the transmission
probability $T(E)$ that is given by
Eq. (\ref{eq:multiT}). The finite voltage,
finite temperature Landauer formula that we use is
\begin{equation}
I(V) = \frac{2e}{h} \int_{-\infty}^{\infty} dE\:T(E)\left(
\frac{1}{\exp[(E-\mu_{s})/kT] + 1} -
\frac{1}{\exp[(E-\mu_{d})/kT]+1} \right)
\end{equation}
where $\mu_s = \epsilon_f + eV/2$ and $\mu_d = \epsilon_f -
eV/2$ and where $\epsilon_f$ is the common Fermi energy of the
leads and $V$ is the applied bias voltage.

The differential conductance at room temperature, $T=293$ K
calculated from the above current for two different choices of
Fermi energy is shown in Fig. \ref{fig5}b.  (It was calculated
using the transmission probability for the dimerized
chain). The solid curve corresponds to a choice of Fermi energy
of -10.2 eV.  Because it lies to the left of the antiresonance
in a region of strong transmission the conductance is strong at
0 V.  It then drops at around 0.2 V when the antiresonance is
crossed.  The dashed curve was calculated using a Fermi energy
of -10.0 eV.  It starts in a region of lower transmission and
thus the antiresonance suppresses the increase in current.
After 0.2 V the large transmission to the left of the
antiresonance is sampled and the current rises sharply.  So in
both cases the antiresonance has served the roll of lowering
the conductance.  It is conceivable to think of utilising more
antiresonances in a narrow energy range to create a more
observable conductance drop.  It should also be pointed out
that the differential conductance was calculated using the
electron transmission evaluated at 0 V.  If one assumes a
linear voltage drop between the leads this approximation is
reasonably valid since the ``active'' molecule is located
roughly in the middle of the bridge between the two leads and
the site energies will not be shifted much by the applied
field.  The coupling elements are assumed not to shift in the
applied field.  With these assumptions, the roots of the
antiresonance condition do not change significantly and so the
location of the antiresonance at -10.08 eV does not shift
appreciably.

\section{conclusions}
In this article we have presented a theoretical study that
suggests that antiresonance phenomena should be observable in
the electrical conductance of molecular wires connected to
metallic nano-contacts. We solved analytically a simple model
that exhibits antiresonances and incorporates for the first
time the effects of the non-orthogonality of tight binding
atomic orbitals on different atoms, an important feature of
all molecular systems. The non-orthogonality was treated
exactly by defining a new energy dependent Hamiltonian
operator and corresponding eigenvector, and then expressing
them in an orthonormal basis of a new Hilbert space. This
method was both simple and very general and should be useful
for treating a wide variety of problems which are best defined
in a non-orthogonal basis.  The Lippmann-Schwinger equation
for the transmission through the molecular wire was solved in
this new representation and an analytic description of the
antiresonances was obtained.  In our model the molecular
antiresonances occur due to two different mechanisms. One of
these is interference between the contributions of different
molecular orbitals as the electron propagates through the
molecule. The other is the vanishing of the effective hopping
matrix element between a pair of atomic orbitals that is due
to the non-orthogonality of those orbitals. In both cases
taking the non-orthogonality into account exactly is necessary
to obtain reliable results.  The antiresonance condition that
we derive and that determines the energies where the
transmission is zero, only depends on the molecular Green's
function and on the energy dependent interaction energies of
the molecular states with the leads.

We have shown that this simple analytically solvable model has
predictive power for more complex systems by performing
detailed numerical calculations.  The first of these
calculations was for a system on which conductance measurements
are not feasible because of the insulating nature of the
leads. However it demonstrated how a molecular wire with a
single $\pi$ state can generate a transmission antiresonance
due to the cancellation of the effective coupling between the
lead and the molecule as a consequence of the mutual
non-orthogonality of atomic orbitals.  The second calculation
was for a system in which molecular antiresonances should in
principle be accessible to experiment: a molecular wire
bridging a break junction between two gold nanocontacts. In it
we have suggested the use of $\pi$ conjugated chains to act as
mode filters. The use of these filters could be useful in most
molecular wire systems where limiting the number of propagating
modes to just one would be advantageous. For filters connected
to an ``active'' molecule it is possible to create an
antiresonance near the Fermi energy of the metal contacts.  The
antiresonance was predicted to manifest itself by producing a
drop in the differential conductance.  In both cases the
location of the antiresonance found in the numerical
simulations was in agreement with the prediction of our simple
analytic model.

We thank R. Akis for helpful correspondence regarding
antiresonances in semiconductor stub tuners.  This work has
been supported by NSERC.

\section{Appendix:  Green's function for Ideal Leads}
The ideal 1D leads are treated using TBA, where the site energy
is $\alpha$, and the nearest neighbour hopping energy is
$\beta$.  The overlap between nearest neighbour lead sites is
$\omega$ and is used to define an energy dependent hopping
parameter $\beta^E = \beta - E \omega$ for an electron with
energy $E$.  This allows us to use an effective orthonormal
basis.  The reduced wavevector, $y$, of a propagating electron
can be found from the equation $\epsilon(y) = \alpha + 2 \beta^E
\cos(y)$.  The leads are semi-infinite.  Because of this the
boundary conditions placed on the wavefunction are such that it
is zero on site 0.  Thus a valid choice for an incident
electron state is a linear combination of a forward and
backward propagating Bloch state, given by,
\begin{equation}
|\Phi_{o}(y)\rangle = \sum_{n=1}^{\infty} (\Phi_{o}(y))_{n} |n
\rangle = \sum_{n=1}^{\infty}
\frac{1}{\sqrt{2N}}(e^{iyn}-e^{-iyn})|n\rangle
\end{equation}

The matrix element for the free propagator of the lead between
sites $n$ and $n'$ is given by,
\begin{equation}
(G_{o}^{R})_{n,n'} = \sum_{y}
\frac{(\Phi_{o}(y))_{n}(\Phi_{o}(y))_{n'}^{*} }{E -
\epsilon(y) + i\delta }
\end{equation}
Using the above expression for $(\Phi_{o}(y))_{n}$, the matrix
element becomes,
\begin{equation}
(G_{o}(E))_{n,n'}= \frac{1}{2N}\sum_{y}
\frac{e^{iy(n-n')}+e^{iy(n'-n)}-e^{iy(n+n')}-e^{iy(-n-n')}}
{E-\epsilon(y)+i\delta}
\end{equation}
As $N$ becomes large the above summation goes over to an
integral which is given by,
\begin{equation}
(G_{o}^{R})_{n,n'} = \frac{1}{2Na}\frac{L}{2\pi}
\int_{-\pi}^{\pi}
\frac{e^{iy(n-n')}+e^{iy(n'-n)}-e^{iy(n+n')}-e^{iy(-n-n')}}{E-
\epsilon(y)+i\delta}\:dy
\end{equation}
where $L=Na$ and
$a$ is the lattice parameter of the chain.  Only the matrix
element on the first site in the lead is needed, since this is
the only site which is coupled to the molecule, so $n$ and
$n'$ are set equal to 1.  Substituting the expression for
$\epsilon(y)$, the following integral is arrived at,
\begin{equation}
(G_{o}^{R})_{1,1} =
\frac{1}{8\pi\beta^E}
\int_{-\pi}^{\pi}\frac{2(1-e^{2iy})}{\frac{E-\alpha}{2\beta^E
}+
\frac{i\delta}{2\beta^E}-\cos{y}
}\:dy
\end{equation}
This integral can be evaluated by performing contour
integration, and the result is,
\begin{equation}
(G_{o}^{R})_{1,1} =
\frac{1}{2\beta^E}\frac{(1-e^{i2y_{o}})}{\sin{y_{o}}}
\end{equation}
where $y_{o}$ satisfies the condition
$\frac{(E-\alpha)}{2\beta^E} - \cos{y_{o}} = 0$.


\begin{references}
\bibitem{LSortho} Antiresonances in the electron transfer
between donors and acceptors in molecules have also been
studied using the LS approach\cite{Ratner90} however the
effects of the non-orthogonality of the tight binding basis
that we consider here were not included in that work.
\bibitem{Lead} In the present calculations, intra-cell and
inter-cell atomic orbital overlap on the leads has not been set
to zero.  The overlap between the unit cell adjacent to the
molecule and the molecule has also been taken into account,
which is an extension of previous work.\cite{Ember98}
\end{references}
\end{document}